\documentclass[12pt]{article}
\usepackage{eurosym}
\usepackage{epsfig}
\usepackage[centertags]{amsmath}
\usepackage{amsfonts}
\usepackage{amssymb}
\usepackage[english]{babel}
\usepackage{amsmath, amsthm, slashed,url}
\usepackage{color}

\begin{document}

\title{\textbf{Zero energy mode for an electron in graphene in a
perpendicular magnetic field with constant asymptotics}}
\author{ Juan Sebasti\'{a}n Ardenghi$^{\dag \ddag }$\thinspace, Alfredo Juan%
$^{\dag \ddag }$\thinspace, Valeria Orazi$^{\dag \ddag }$\thinspace, Lucas~Sourrouille$^{\dag \ddag }${\normalsize 
\textit{$^{a}$}}\thinspace , \\
$^{\dag }$Departamento de F\'{\i}sica, Universidad Nacional del Sur, Av.
Alem 1253,\\
B8000CPB, Bah\'{\i}a Blanca, Argentina \\
$^{\ddag }$Instituto de F\'{\i}sica del Sur (IFISUR, UNS-CONICET), Av. Alem
1253,\\
B8000CPB, Bah\'{\i}a Blanca, Argentina\\
{\normalsize \textit{$^{a}$}}{\footnotesize sourrou@df.uba.ar} }
\maketitle


\abstract{We study the influence of a perpendicular magnetic field with the asymptotics $B(r\to \infty)= B_0$ in a 
electrons in graphene. It is shown that the zero-energy  solutions can exist only for one pseudospin direction, depending on the
sign of the magnetic field in the infinite boundary. This, 
shows that zero-energy level is robust with respect to possible inhomogeneities of the magnetic field. In  addition, we show that 
the number of the states with zero energy for one pseudospin projection is infinity. These results should be
useful in the study of ripples which cause a scattering of Dirac particles in slowly decreasing magnetic fields where the 
asymptotic states is easy to define. }




\section{Introduction}

The energy spectrum of fermions coupled to gauge fields presents an
important property, the existence of zero-energy modes equally shared by
electrons and holes. This is a consequence of one of the most important
theorems of modern mathematics, the Atiyah-Singer index theorem \cite{1,2}.
This theorem has important applications in quantum field and superstring
theories \cite{3,4}. It, also, has important consequences on graphene \cite%
{5,6} where the anomalous (half-integer) quantum Hall effect in single-layer
graphene is a consequence of the Atiyah-Singer theorem. In addition, a
similar statement has been proven for the anomalous quantum Hall effect in
graphene bilayer \cite{7,8}. \newline
The existence of the zero-energy states for fermions in inhomogeneous
magnetic fields in two dimensions was demonstrated explicitly by Aharonov
and Casher \cite{ahronov}. They, was able to show that the ground state is
exactly calculable and possesses a degeneracy related to the total magnetic flux.
Specifically, they suppose the case of an infinite sample with the magnetic
flux $\Phi $ localized in a restricted region and showed that zero-energy
solutions can exist only for one spin direction, depending on the sign of
the total flux. Also, they showed that the number of the states with zero
energy for one spin projection is equal to $N$, where $N$ is the integer
part of $\frac{\Phi }{2\pi }$. 

In this note we generalize the idea of Aharonov and Casher and study
electrons in graphene under the influence of perpendicular magnetic field
with the asymptotics $B(r\rightarrow \infty )=B_{0}$, where $B_{0}$ is a real constant. Particularly, we show
that the zero-energy solutions can exist only for one pseudospin direction,
that in graphene implies that the wave function of electrons are not
vanishing only for one sublattice basis and in turn this depends on the sign
of the magnetic field in the boundary. In addition, we show that the number
of the states with zero energy for one pseudospin projection is infinity.
These results are importnat in nanophysics, because it is expected that
graphene will serve a base for new optoelectronic devices (\cite{5}, \cite%
{zhang}). In particular, in order to control the motion of electrons, the
use of non-homogeneous magnetic fields has been study, through the deposit
of nanostructured ferromagnets (\cite{pet1}, \cite{reij}, \cite{reij2}). In
recent works (\cite{jsa1}, \cite{jsa2}, \cite{jsa3}, \cite{jsa4} and \cite%
{jsa5}) , the magnetic properties of graphene under constant perpendicular
magnetic field taking into account the Zeeman effect and Rashba spin-orbit
coupling has been studied, showing that zero energy states alter drastically
the magnetic oscillations and the quantum Hall effect. In order to extend to
the variable magnetic fields, we consider a constant asymptotics for the
magnetic field that model a magnetic antidot, where $B$ is zero in a
circular region and nonzero outside this region which can be achieved by
having a magnetic vortex piercing the graphene layer \cite{reij2}.\ This
work will be organized as follow: In section II, the simplified model for
graphene in a constant asymptotic non-vanishing magnetic field is
introduced. In section III, the solutions ar computed and discussed- The
principal findings of this paper are highlighted in the conclusion.

\section{The model}

Let us start by considering electrons in graphene in the long wavelength
approximation, where the $(2+1)$-dimensional Dirac-Weyl model whose
Hamiltonian is described by 
\begin{equation}
H=(\sigma _{x}p_{x}+\sigma _{y}p_{y})
\end{equation}%
Here, the $\sigma _{i}$ $(i=x,y)$ are 2$\times $2 Pauli matrices acting in
the sublattice basis, i.e. 
\begin{equation*}
\sigma _{x}=\left( 
\begin{array}{cc}
0 & 1 \\ 
1 & 0%
\end{array}%
\right) \,,\;\;\;\;\;\ \sigma _{y}=\left( 
\begin{array}{cc}
0 & -i \\ 
i & 0%
\end{array}%
\right) 
\end{equation*}%
$p_{i}=-i\partial _{i}$ is the two-dimensional momentum operator and for
simplicity we have considered that the Fermi velocity $v_{F}=1$. The
massless Dirac-Weyl equation in $(2+1)$ dimensions is 
\begin{equation}
\sigma ^{i}p_{i}\Psi (x,y,t)=i\partial _{t}\Psi (x,y,t)  \label{eq1}
\end{equation}%
Here, $\Psi (x,y,t)$ is the two-component spinor 
\begin{equation}
\Psi =(\psi _{A},\psi _{B})^{T}
\end{equation}%
here $\psi _{a}$ and $\psi _{b}$ represent the envelope functions associated
with the probability amplitudes in each sublattice basis. Since, we are
interested in stationary states, it is natural to propose a solution of the
form 
\begin{equation*}
\Phi (x,y,t)=e^{-iEt}\Psi (x,y)\;
\end{equation*}%
then, the time-independent Dirac-Weyl equation is 
\begin{equation}
\sigma ^{i}p_{i}\Psi (x,y)=E\Psi (x,y)  \label{1dw}
\end{equation}%
In the presences of a perpendicular magnetic field to the $(x,y)$-plane, we
replace the momentum operator $p_{i}$ by the covariant derivative, defined
as $D_{i}=-i\partial _{i}+A_{i}$ $(i=x,y)$, where $A_{i}$ are components of
the vector potential, such that 
\begin{equation}
B(x,y)=\partial _{x}A_{y}-\partial _{y}A_{x}\;  \label{mag}
\end{equation}%
where $B(x,y)$ is the transversal magnetic field. Thus, the equation (\ref%
{1dw}) becomes, 
\begin{equation}
\sigma ^{i}D_{i}\Psi (x,y)=E\Psi (x,y)  \label{2dw}
\end{equation}%
We can develop this equation to get 
\begin{equation}
\left( 
\begin{array}{cc}
0 & D_{1}-iD_{2} \\ 
D_{1}+iD_{2} & 0%
\end{array}%
\right) \left( 
\begin{array}{c}
\psi _{A} \\ 
\psi _{B}%
\end{array}%
\right) =E\left( 
\begin{array}{c}
\psi _{A} \\ 
\psi _{B}%
\end{array}%
\right)   \label{3dw}
\end{equation}%
where $\psi _{A}$ and $\psi _{B}$ are the sublattice components of the
spinor $\Psi $ (i.e. $\Psi =(\psi _{a},\psi _{b})^{T}$). From this equation
we can write the two coupled equations for the components $\psi _{A}$ and $%
\psi _{B}$ 
\begin{equation}
\lbrack D_{1}-iD_{2}]\psi _{B}=E\psi _{A}  \label{eqm1}
\end{equation}%
\begin{equation}
\lbrack D_{1}+iD_{2}]\psi _{A}=E\psi _{B}  \label{eqm2}
\end{equation}%
Here, we are interested in the zero energy modes. This solution may be
constructed explicitly following, the work done by Aharonov and Casher \cite%
{ahronov}. Thus, we can introduce a scalar potential $\lambda (x,y)$ such
that, 
\begin{equation}
A_{x}=-\partial _{y}\lambda \,,\;\;\;\;\;\ A_{y}=\partial _{x}\lambda 
\label{gau}
\end{equation}%
and due to the equation (\ref{mag}), 
\begin{equation*}
B=\partial _{x}^{2}\lambda +\partial _{y}^{2}\lambda 
\end{equation*}%
Then, the equations (\ref{eqm1}) and (\ref{eqm2}), for the energy zero case,
can be written in the form 
\begin{equation}
\Big[\lbrack -i\partial _{x}-\partial _{y}]+(-\partial _{y}\lambda
-i\partial _{x}\lambda )\Big]\psi _{B}=0  \label{eqm1.1}
\end{equation}%
\begin{equation}
\Big[\lbrack -i\partial _{x}+\partial _{y}]+(-\partial _{y}\lambda
+i\partial _{x}\lambda )\Big]\psi _{A}=0  \label{eqm2.1}
\end{equation}%
The potential $\lambda (x,y)$ can be excluded by the substitution 
\begin{equation}
\psi _{A,B}=f_{A,B}e^{\gamma \lambda }  \label{asy}
\end{equation}%
where $\gamma =1$ and $-1$ for $\psi _{A}$ and $\psi _{B}$ respectively.
Then the equations (\ref{eqm1.1}) and (\ref{eqm2.1}) transforms into the
equation%
\begin{equation}
\lbrack -i\partial _{x}+\gamma \partial _{y}]f_{A,B}=0  \label{f1}
\end{equation}%
Thus, $f_{A}$ and $f_{B}$ are analytic and complex conjugated analytic
entire functions of $z=ix+y$, respectively. \newline
For a transversal magnetic field $B(x,y)$ with the asymptotics $%
B(r\rightarrow \infty )=0$ the problem was addressed by Aharonov and Casher
in Ref.\cite{ahronov}, and leads to zero-energy solutions exist only for one
spin direction, depending on the sign of the total magnetic flux. Also they
proved that the number of the states with zero energy for one ospin
projection is equal to $N$, where $N$ is the the closest integer to the
total flux in units of the flux quantum. \newline
In this letter we are interested in magnetic fields with the asymptotics $%
B(r\rightarrow \infty )=B_{0}$, where $B_{0}$ is a constant real number
different from zero. Hence, as $r\rightarrow \infty $ the equation (\ref{mag}%
) becomes, 
\begin{equation}
B_{0}=\partial _{x}A_{y}-\partial _{y}A_{x}\;  \label{mag1}
\end{equation}%
which can be solved by 
\begin{equation}
A_{x}=-\frac{B_{0}}{2}y\,,\;\;\;\;\;\ A_{y}=\frac{B_{0}}{2}x  \label{gau1}
\end{equation}%
which coincides with the symmetric gauge. To proceed, we can obtain an
expression for $\lambda (x,y)$ by using (\ref{gau}) and (\ref{gau1}), 
\begin{equation*}
\lambda (r\rightarrow \infty )=\frac{B_{0}}{4}(x^{2}+y^{2})=\frac{B_{0}}{4}%
r^{2}
\end{equation*}%
Then, the solution of Eq. (\ref{asy}) at large distances has the
asymptotics, with 
\begin{equation*}
\psi _{A,B}=f_{A,B}e^{\gamma \frac{B_{0}}{4}r^{2}}
\end{equation*}%
Since the entire function $f_{A,B}(z)$ cannot go to zero in all directions
at infinity, $\psi _{i}$ can be normalizable only assuming that $\gamma 
\frac{B_{0}}{4}<0$; that is, zero-energy solutions can exist only for one
spin direction, depending on the sign of the magnetic field in the boundary.
Here, it is important to remark that our result remains robust with respect to possible inhomogeneities of
the magnetic field in a finite region of the plane. In other words that means that the spin polarization, depends only on 
asymptotic behavior of the magnetic field. This statement is important for real graphene since the effective magnetic field there
should be inhomogeneous due to the effect of so-called ripples \cite{5}, \cite{6}.
\\
To conclude this section we can count how many independent solutions of Eq. (\ref{f1}) we have. As a basis, we can choose just 
polynoms searching
the solutions of the form
\begin{eqnarray} 
\psi_{b} = z^j e^{ -\frac{B_0}{4} r^2}
\label{sol1}
\end{eqnarray}
(to be specific, we consider the case $B_0 > 0$), where j = 0, 1, 2, . . . Since $e^{ -\frac{B_0}{4} r^2}$ decays faster than any 
power of $z$. One can easily see from Eq. (\ref{sol1}) that the solution is integrable with the square for any value of $j$.
Therefore, the number of the states with zero energy for one spin projection is infinity, and there are no such solutions for 
another spin projection.

\section{Results and discussion}
Let us, now, discuss the applications of our result to graphene systems.
In graphene, the spin polarization is a pseudospin polarization,
that is reflected in a non-vanishing envelope function only for the $A$ or $B
$ sublattice of the Bravais lattice of graphene depending on the sign of $%
B_{0}$. Indeed, graphene has a degeneracy related to the Dirac points in
the reciprocal lattice that introduces the valley index $\tau =\pm 1$, where 
$+1$ is for the $K$ valley and $\,-1$ for the $K^{\prime }$ valley, where $K$
and $K^{\prime }$ are two inequivalent Dirac points in $k$-space (\cite{jsa6}%
, \cite{lian}). The Hamiltonian can be written as $H=\tau _{z}\otimes
(\sigma _{x}p_{x}+\tau \sigma _{y}p_{y})$ where~$\tau _{z}$ is the $\sigma
_{z}$ operator in the valley subspace. Following this, the wavefunction can
be written as $\psi _{A,B}=f_{A,B}e^{\tau \gamma \lambda }$. Then, for $%
B_{0}>0\,\ $in the $K$ valley only the $B$ sublattice component is
non-vanishing and in the $K^{\prime }$ valley only the $A$ sublattice is
non-vanishing, that is, the case of zero energy is special since its amplitude
is nonzero only in one of the sublattices, namely, at $B$ sites for $K$ and $%
A$ sites for $K^{\prime }$. Due to the degeneracy of the Hamiltonian and
when no other external mechanism is consider to break the valley symmetry of
the Hamiltonian, the zero mode solution consists in a superposition in the
pseudospin space

\begin{eqnarray}
\left\vert \Psi _{+}\right\rangle  &=&\frac{1}{\sqrt{2}}e^{-\frac{B_{0}}{4}%
r^{2}}(f_{BK}\left\vert B\right\rangle \left\vert K\right\rangle
+f_{AK^{\prime }}\left\vert A\right\rangle \left\vert K^{\prime
}\right\rangle )  \label{r1} \\
\left\vert \Psi _{-}\right\rangle  &=&\frac{1}{\sqrt{2}}e^{\frac{B_{0}}{4}%
r^{2}}(f_{AK}\left\vert A\right\rangle \left\vert K\right\rangle
+f_{BK^{\prime }}\left\vert B\right\rangle \left\vert K^{\prime
}\right\rangle )  \notag
\end{eqnarray}%
where $\left\vert \Psi _{+}\right\rangle $ is for $B_{0}>0$ and $\left\vert
\Psi _{-}\right\rangle $ is for $B_{0}<0$. The entanglement between the
valley and pseudospin is manifest and can be tuned with the magnetic antidot
by alternating the asymptotic behavior of the magnetic field. In turn, $%
\left\langle \Psi _{+}\mid \Psi _{-}\right\rangle =0$ and no transition
between zero mode solutions is possible. In order to obtain the number of
independent solutions, the differential equation for $f$ \ is in eq.(\ref{f1}%
) only for the $K$ valley, although by symmetry, $f_{AK^{\prime }}=f_{BK}=%
\overline{z}^{n}$ ,\ \ $f_{BK^{\prime }}=f_{AK}=z^{n}$ can be obtained. For
single valuedness, the function $f$ \ has to be a polynomial, which implies
that $f$ must be of degree $n$, where $n$ is determined by the total flux $%
\Phi =\int BdA=(n+\epsilon )\phi _{0}$ where $\phi _{0}=\hbar /e$ is the
quantum of magnetic flux and $0<\epsilon <1$. Then the Hamiltonian for each
valley has therefore $n$ zero-modes independently of the shape of the
magnetic field in the interior.

In this sense, graphene with constant asymptotics magnetic field has a
chiral symmetry in the massless limit of the Dirac equation, except for the
zero modes. Should be stressed that the symmetry in the Hamiltonian in
graphene is a manifestation of supersymmetry \cite{wit} in quantum
mechanics, which appears as a quantum Hall spectrum in multilayer graphene 
\cite{eza} and is robust for arbitrary magnetic fields. In particular, for
random magnetic fields, as it can occur with ripples \cite{voz}, Dirac
particles scatters slowly decreasing magnetic fields whose asymptotics can
be positive or negative randomly and the scattering states can be
represented by eq.(\ref{r1}) with arbitrary space-dependent $f$
coefficients. Consequently, while a random on-site disorder potential gives
only intravalley mixing within either the $K$ and $K^{\prime }$ valleys,
random hopping can cause intervalley mixing. Nevertheless, for general
hopping disorder potential $U$ (see eq.(7) of \cite{pereira}), $\left\langle
\Psi _{+}\right\vert U\left\vert \Psi _{-}\right\rangle =0$, no transition
between asymptotic zero modes is allowed to first order in the random
potential and intervalley mixing is expected for $n>0$ modes or for $n=0~$at
second order in the random potential.

\section{Conclusions}

In summary, we have generalized the result found in Ref.\cite{ahronov} for
the cases in which the magnetic field is different from zero as $%
r\rightarrow \infty $. In particular, we showed that the zero-energy
solutions can exist only for one pseudospin direction in graphene, depending
on the sign of the magnetic field on the infinite edge. Thus, the
zero-energy level is protected by the value of the magnetic field in
boundary, which show that it is robust with respect to possible
inhomogeneities of the magnetic field. Finally, we have considered the valley degrees
of freedom, we habe found that zero energy mode has two orthogonal entangled
solution in the valley and pseudospin states and that can be tuned with the
magnetic field asymptotics.

\bigskip 

\textbf{Acknowledgements} \newline
This paper was partially supported by grants of CONICET (Argentina National
Research Council) and Universidad Nacional del Sur (UNS)and by ANPCyT
through PICT 1770, and PIP-CONICET Nos. 114-200901-00272 and
114-200901-00068 research grants, as well as by SGCyT-UNS., J. S. A. and L.
S. are members of CONICET

\end{document}